\documentclass[aps,prb,twocolumn, superscriptaddress]{revtex4-1}
\usepackage{color}
\usepackage{xcolor}
\usepackage{amsmath}
\usepackage{amssymb}
\usepackage{graphicx}
\usepackage{lipsum}
\usepackage{subfigure}	
\begin{document}
\title{Role of the compensating current in the weak Josphson coupling regime: An extended study on excitonic Josephson junctions} 

\author{Ya-Fen Hsu}
\email[E-mail address: ]{yafen.hsu.jane@gmail.com}
\affiliation{Physics Division, National Center for Theoretical Science, Hsinchu, 30013, Taiwan}
\affiliation{Department of Electrophysics, National Chiao Tung University, Hsinchu 300, Taiwan}

\author{Jung-Jung Su}\email[E-mail address: ]{jungjsu@nctu.edu.tw}
\affiliation{Department of Electrophysics, National Chiao Tung University, Hsinchu 300, Taiwan}
\date{\today}
\begin{abstract}
Huang's experiment [Phys. Rev. Lett. $\bf 109$, 156802 (2012)] 
found, in the quantum Hall bilayer of the Corbino geometry, 
the interlayer tunneling currents at two edges are coupled to each other 
and one of two tunneling currents is referred to as the compensating current of the other.  
The recent theoretical work [arXiv:2006.15329] has explained this exotic coupling phenomenon 
as a result of excitonic Josephson effect induced by interlayer tunneling current. 
In this paper, we study the same setup---excitonic Josephson junction--- but in the weak Josephson coupling regime, 
which occurs for large junction length. 
Interestingly, we find the compensating current drives the other edge to undergo a nonequilibrium phase transition 
from a superfluid to resistive state, which is signaled by an abrupt jump of the critical tunneling current. 
We also identify the critical exponent and furthermore offer more experimental prediction.         
\end{abstract}

\maketitle


\section{Introduction}
Josephson effect is particularly attractive to condensed matter researchers because 
it serves as the striking manifestation of coherent condensation and the promising candidate for quantum technology.  
The unrelenting and strong attention has been received recently in optically-excited exciton or exciton-polariton cold gases
\cite{Carusotto,Shelykh,Lagoudakis,Rontani:PRL2010,Abbarchi,Adiyatullin,Caputo} 
and graphene electron-hole bilayer exciton\cite{Zenker,Apinyan}. 
However, being the best platform to achieve the exciton condensation, the quantum Hall bilayer
\cite{Girvintextbook,Eisenstein:ARCRP2014,Spielman:PRL2000,Kellogg:PRL2002,Tutuc:PRL2003,Eisenstein:Nature2004,Kellogg:PRL2004,Tutuc:PRL2004,Wiersma:PRL2004,Tiemann:NJP2008,Su:NatPhys2008,Misra:PRB2008,Tieleman,Yoon,Fink:PRL2011,Nandi:Nature2012,Nandi:PRB2013,Cipri,Zhang:PRL2016,Barkeshli,Sodemann,Barkeshli,Eisenstein:PRL2019,Zhu:PRB2019,Zhang:PRL2020} remains not studied extensively in the land of Josephson effect. 
Actually, the search for Josephson effect in quantum Hall bilayer ever arouse intense interest 
since the observation of Josephson-like tunneling\cite{Girvin2,Wen4}, 
in which the interlayer voltage abruptly increases once exceeding a critical tunneling current
\cite{Spielman:PRL2000,Spielman:PRL2000,Eisenstein:Nature2004,Tiemann:NJP2008,Misra:PRB2008,Yoon,Nandi:PRB2013}. 
However, to the end, the Josephson-like tunneling is attributed to 
a mixture of coherent and incoherent interlayer tunneling\cite{Joglekar,Rossi,JJSu:PRB2010} 
instead of the ``real" Josephson effect. 
Once exceeding a critical current, the incoherent tunneling dominates over the coherent one.
   
The scattering approach by solving the Bogolubov-de Gennes Hamiltonian\cite{Titov,Dolcini,Peotta} 
is the standard one to explore the Josephson effect 
but it is difficult to access in the context of quantum Hall bilayer. 
In our previous works\cite{YFHsu:SR2015,YFHsu:NJP2018}, we therefore turn to a new method within the frame of pseudospin dynamics 
based on the idea that the layers can be treated as pseudospin quantum degrees of freedom\cite{JJSu:PRB2010,Moon:PRB1995,Burkov:PRB2002}. 
We firstly employ this new method to study the exciton-condensate/exciton-condensate (EC/EC)\cite{YFHsu:SR2015} 
and exciton-condensate/normal-barrier/exciton-condensate (EC/N/EC) junctions\cite{YFHsu:NJP2018} 
with a constant relative phase between two ECs that is generated by perpendicular electric field\cite{Wen:EPL1996}.   
We found that excitonic Josephson effect occurs only when $d_J\leq\xi$, 
where $d_J$ and $\xi$ are barrier length and correlation length\cite{YFHsu:SR2015,YFHsu:NJP2018}. 
When $d_J>\xi$, a new transport mechanism, namely, tunneling-assisted Andreev reflection occurs at a single N/EC interface\cite{YFHsu:NJP2018}.   
While the excitonic Josephson effect gives rise to novel fractional solitons\cite{YFHsu:SR2015}, 
the new mechanism leads to a half portion of fractional solitons\cite{YFHsu:NJP2018}.
Notably, these new types of solitons have potential to improve the stability and efficiency of quantum logic circuits\cite{Pegrum}. 
We next study another setup suggested to have a relative phase by externally applying interlayer tunneling current\cite{Park:PRB2006}.

\begin{figure}  
\centering
\includegraphics[width=7.7cm]{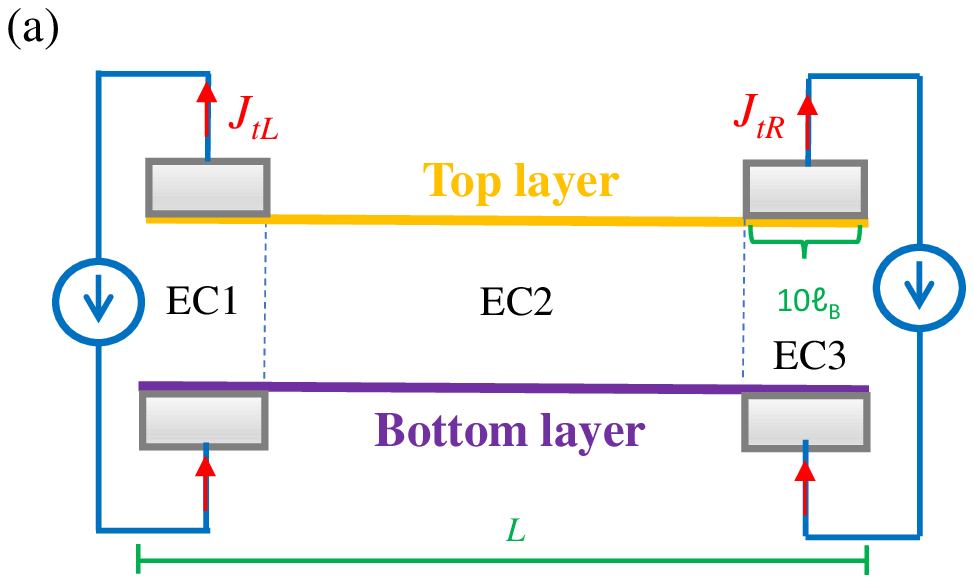}
\includegraphics[width=7.1cm]{EXJETCLFig1b.eps}
\includegraphics[width=7.1cm]{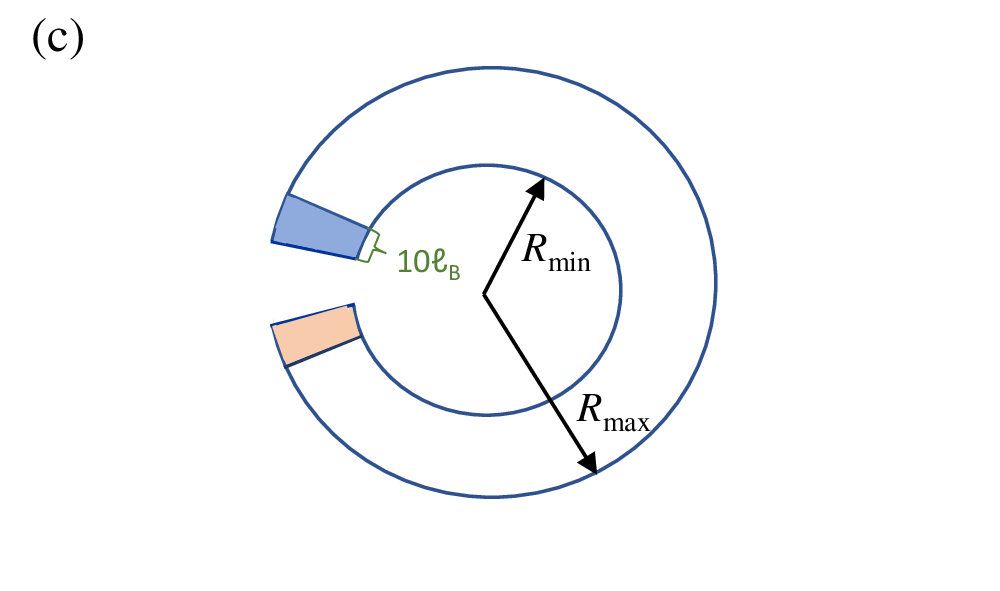}
\vspace{-0.5cm}
\caption{(color online) (a) Schematic layout of an excitonic Josephson junction induced by interlayer tunneling current. 
The relative phases between three condensate regions: EC1,EC2 and EC3, are generated 
by externally applying tunneling currents $J_{tL}$ and $J_{tR}$. 
$\ell_B$ and $L$ denote the magnetic and junction length.  
(b) The calculated phase distributions for parallel polarity $\oplus$ ($J_{tL}=J_{tR}$) 
and anti-parallel polarity $\ominus$ ($J_{tL}=-J_{tR}$) with $L=12\lambda$.  
The green (black) and pink (grey) lines correspond to the parallel and anti-parallel polarity, respectively.
The employed values of $J_{tR}$ are 5,10,15,20,25 $J_{t0}$ and with increasing $J_{tR}$, 
the phase $\phi$ departs from the $x$ axis. 
The length unit $\lambda$ and the current unit $J_{t0}$ are given later in Sec. \ref{unit}. 
Such a long junction is similar to two weakly coupled exciton-condensate/exciton-condensate (EC/EC) junctions. 
The cross is the breakpoint between two EC/EC junctions and it is located where the Josephson current $J_s$ approaches zero. 
The left (right) part of the bulk combines with the left (right) edge forming an EC/EC junction. 
$L_{\rm{eff}}$ denotes the effective junction length of the right EC/EC junction.  
(c) Schematic layout of a Corbino-geometry excitonic Josephson junction. 
The two tunneling currents $J_{tL}$ and $J_{tR}$ are exerted on 
the orange (lower) and blue (upper) shadow zones. 
$R_{\rm{min}}$ and $R_{\rm{max}}$ are the minimum and maximum radius.     
}
\label{Fig1}
\end{figure}

Inspired by Huang's experiment\cite{Huang:PRL2012}, we consider the setup of interlayer tunneling currents  
exerted on two edges of quantum Hall bilayer as shown in Fig. \ref{Fig1}(a). 
The tunneling currents ($J_{tL}$,$J_{tR}$) twist the condensate phases of two edges 
so as to create the relative phases between three condensates: EC1, EC2, and EC3.
Such structure is regarded as two condensates (EC1 and EC3) sandwiched by a superfluid barrier (EC2), 
which is a type of excitonic Josephson junctions\cite{Golubov}. 
Ref. \cite{YFHsuPRL} has explored this setup but focuses on the short junction 
whose junction length $L$ is smaller than Josephson length $\lambda$\cite{note1}. 
Its results demonstrated that the exotic coupling phenomenon of edge tunneling currents observed by Huang {\it et al}\cite{Huang:PRL2012} 
is originated from excitonic Josephson effect and Huang's experiment is a very robust evidence 
for quantum Hall bilayer exciton condensation. 

In this paper, we turn our attention to the opposite case---the long junction of $L\sim10\lambda$, which corresponds to 
the typical quantum Hall bilayer\cite{Hyart:PRB2011,Spielman:PRL2000}. 
Our calculation of the condensate phase [see Fig. \ref{Fig1}(b)] reflects that 
the Josephson current is essentially negligible in the bulk since the phase goes to zero and becomes flat there\cite{Js}.         
Because the two edges are weakly Josephson coupled, the long junction can be approximated as  
two independent EC/EC junctions with the boundary between them occurring 
where Josephson current $J_s$ goes to zero [see the inset of Fig. \ref{Fig1}(b)].  
It is therefore highly desirable that the long junction can display entirely different properties 
from the short junction in which two edges are strongly Josephson coupled\cite{YFHsuPRL}. 

It turns out that the long junction indeed exhibits an unique property: 
one edge undergoes a nonequilibrium phase transition\cite{Nakamura:PRL2012,Matsumoto:PRD2018} 
with increasing the tunneling current at the other edge (i.e., the compensating current). 
During this phase transition, the critical tunneling current of the edge sharply falls 
and the corresponding critical exponent is identified as $\gamma\sim0.5$.  
Since the Josephson coupling is weak, we wonder why the compensating current 
can influence the other edge so largely? 
According to our analysis, this is because the compensating current 
reduces the effective junction length of the constituent EC/EC junction on the opposite side. 
We furthermore calculate the magnetic field induced by Josephson current (denoted by $B_J$)
for the Corbino-geometry excitonic Josephson junction shown in Fig. \ref{Fig1}(c). 
We find the length reduction effect is revealed 
by the crossover of the $B_J$ versus $\Delta J_t$ curve 
into the short junction regime\cite{YFHsuPRL} (a linear one) with increasing the compensating current, where $\Delta J_t=J_{tR}-J_{tL}$.   
The induced magnetic field is estimated at $\sim100$pT that is large enough to 
be detected by the scanning superconducting interference device (SQUID). 
In the main body of this paper, we show the results of the rectangle-shaped junction in Figs. 3-7 
while that of the Corbino-geometry junction in Figs. 8-9.

\section{Model and method}
Burkov and MacDonald treated two layers of the quantum Hall bilayer as pseudospin quantum degrees of freedom 
and accordingly deduced a lattice model Hamiltonian\cite{Burkov:PRB2002}: 
\begin{align}
H=\frac{1}{2}\sum_{ij}(2H_{ij}-&F^{\rm{intra}}_{i,j})S^z_iS^z_j-F^{\rm{inter}}_{i,j}(S^x_iS^x_j+S^y_iS^y_j),\notag\\
&\vec{S}_i=\frac{1}{2}\sum_{\sigma,\sigma'}a^\dagger_{i,\sigma}\vec{\tau}_{\sigma,\sigma'}a_{i,\sigma'}. 
\end{align}   
Here $a^\dagger_{i,\sigma}$($a_{i,\sigma}$) is the Schwinger boson creation (annihilation) operators\cite{Lacroixtextbook} 
where $i$ and $\sigma$ label the site and layer indexes and $\vec{\tau}$ is the Pauli matrix vector. 
The Hartree term $H_{ij}$ describes the direct Coulomb interaction while 
the Fock term $ F^{\rm{intra}}_{i,j}$ ($ F^{\rm{inter}}_{i,j}$)  
serves the intralayer (interlayer) exchange interaction. 
This lattice Hamiltonian possesses eigenstate wave function which can be generally expressed as   
\begin{equation}
	|\Psi\rangle=\prod_{i}\left[\cos\frac{\theta(\vec{X}_i)}{2}c^{\dagger}_{i\uparrow}
	+\sin\frac{\theta(\vec{X}_i)}{2}e^{{\it i}\phi(\vec{X}_i)}c^{\dagger}_{i\downarrow}\right]|0\rangle. \label{wave2}
\end{equation}
The operator $c^{\dagger}_{i\uparrow}$($c^{\dagger}_{i\downarrow}$) creates an electron 
at the lattice site location $\vec{X}_i$ in the top (bottom) layer. 
It is difficult to study the present issue through quantum scattering approach which is based on this wave function 
since we cannot simply write down the explicit forms of $\theta(\vec{X}_i)$ and $\phi(\vec{X}_i)$. 

We therefore request a SU(2) to O(3) mapping and the wave function is transformed into a classical pseudospin\cite{Moon:PRB1995} 
\begin{align}
\vec{m} (\vec{X}_i)& = (m_{\perp} \cos \phi,\ m_{\perp} \sin \phi,\ m_z ), \notag\\
  &m_{\perp}=\sin\theta,\ m_z=\cos\theta. 
\end{align}  
Accordingly, the dynamics of the quantum Hall bilayer can be described 
by the Landau-Lifshitz-Gilbert (LLG) equation\cite{YFHsu:SR2015,YFHsu:NJP2018,JJSu:PRB2010}  
\begin{align}
\frac{d \, \vec{m}}{d\,t}&=\vec{m}\times (2/n \hbar)({\delta E[\vec{m}]}/{ \delta \vec{m} })
-\alpha\left(\vec{m}\times \frac{d\vec{m}}{dt}\right),\notag\\    
E[\vec{m}] &= A_{\rm{unit}} \sum_i \left[ \beta m^2_z
+\frac{\rho_s m^2_{\perp}}{2}|\nabla_{\vec{X}_i} \phi|^2\right.\notag\\ 
& \left.\mbox{ }-\frac{n\Delta_tm_{\perp}}{2}\cos\phi\right], 
  \label{LLG}  
\end{align}
where $A_{\rm{unit}}$ is the area of the unit cell for the pseudospin lattice and $n$ is the pseudospin density. 
The excitonic superfluid loses its coherence after traveling over one correlation length $\xi$ so 
the size of the unit cell is equal to $\xi$, which is estimated at $\sim200$nm\cite{Eastham:PRB2009}. 
In unit of the magnetic length $l_B$, $\xi\sim10l_B$ ($l_B$ has the typical value of $\sim20$nm). 
On the other hand, the energy functional $E[\vec{m}]$ is composed of the capacitive penalty, 
the exchange correlation, and the interlayer tunneling energy, 
which are characterized by the parameters: anisotropic energy $\beta$, pseudospin stiffness $\rho_s$, 
and single-particle tunneling $\Delta_t$, respectively. 
These model parameters is up to which kind of samples we are discussing. 
The second term for the LLG equation is the Gilbert damping which relaxes the energy toward the minimum. 
\vspace{0.1cm} 
\begin{figure} 
\includegraphics[width=6.0cm]{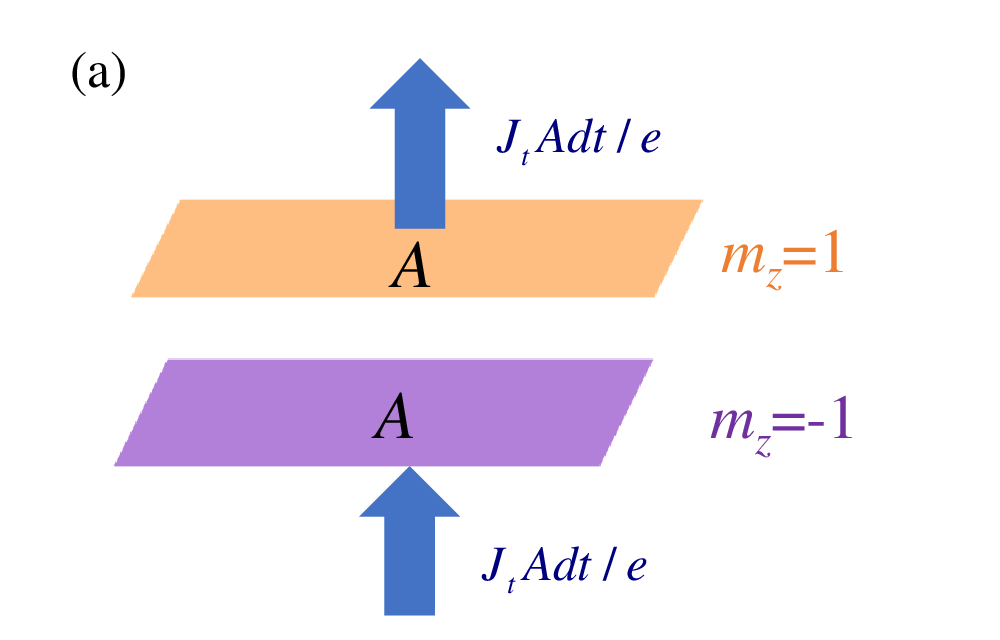}
\centering
\includegraphics[width=7.5cm]{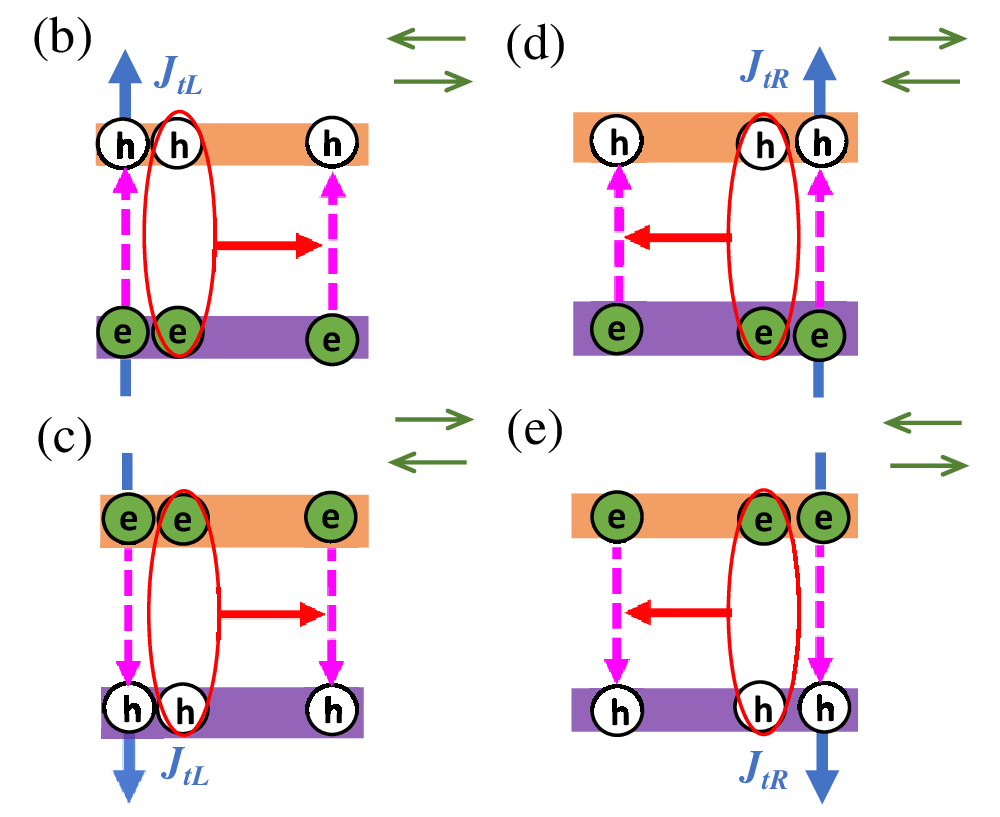}
\caption{(color online) (a) Illustration of the effect of external tunneling current $J_t$. 
Here the top and bottom layers are selected as up pseudospin ($m_z=1$) and down pseudospin ($m_z=-1$).  
The notation $A$ denotes the area that tunneling current passes through.  
Over the time duration $dt$, the electrons number that flows out of the top layer 
or flows into the bottom layer is counted by $J_tAdt/e$. 
(b),(c) and (d),(e) depict the flows of electrons and excitons  
when applying the external tunneling current to the right and left edges, respectively.    
The solid red and dashed pink arrows indicate the direction of exciton flow and single-particle tunneling, respectively.  
The insets at their upper right corner are the individual corresponding counterflow currents. 
For convenience in discussion, we choose $+e$ as the charge of an electron and $e$ is actually a negative amount. 
The current therefore goes along the flow direction of electrons.  
} 
\label{Fig2}
\end{figure}
\subsection{Modeling excitonic Josephson junctions}
\label{method}
The key breakthrough of the present work is to introduce the effect of external tunneling currents. 
When exerting the $+\hat{z}$-direction tunneling current $J_t$ on a area of $A$ over a short duration of $d t$, 
there are electrons as many as $J_tAdt/e$ pouring out of the top layer 
and trickling into the bottom layer simultaneously (see Fig. \ref{Fig2}), 
giving rise to the change of $-2J_tAdt/e$ in the total pseudospin $nAm_z$. 
Under the effect of tunneling current, the $z$-component LLG equation thus can be modified as 
\begin{align}
\frac{dm_z}{dt}=-\frac{2\rho_s}{n\hbar}m^2_{\perp}\nabla^2\phi+\frac{\Delta_t}{\hbar}m_{\perp}\sin\phi
-\frac{2J_t}{ne}+\alpha m^2_{\perp}\frac{d\phi}{dt}. 
\label{MLLG}
\end{align} 
In the rectangle-shaped excitonic Josephson junction as shown in Fig. \ref{Fig1}(a), 
two tunneling current $J_{tL}$ and $J_{tR}$ are applied to two edges over a length as large as one lattice size 10$\ell_B$. 
We can therefore model the junction through setting $J_t$ to 
\begin{align}
J_t&=J_{tL}\Theta(x+L/2)\Theta(L/2-10l_B-x)\notag\\
&+J_{tR}\Theta(L/2-x)\Theta(x-L/2+10l_B).
\end{align} 
Notice we from here on use the continuous varying $x$ instead of the discrete $\mbox{X}_{i}$ for convenience in presentation  
and $\Theta(x)$ is the Heaviside step function. 
The origin $x=0$ is defined to be located at the center of the junction. 
After evolving with time, we ultimately acquire the static solutions for $\phi$, $m_{\perp}$, and $m_z$ 
that specify the pseudospin orientation.  
The counterflow Josephson current is furthermore calculated by 
\begin{equation}
J_s=e\rho_s\nabla\phi/\hbar.   
\label{Js}
\end{equation}

The physical picture for the effect of external tunneling currents can be depicted through Figs. \ref{Fig2}(b)-(e).   
When applying the $+\hat{z}$-direction tunneling current to the left edge [see Fig. \ref{Fig2}(b)], 
holes and electrons are injected into the top and bottom layer from the left side, respectively. 
The electrons can flow into the top layer to annihilate holes via single-particle tunneling $\Delta_t$ or 
combine with holes to form excitons $|h\uparrow;e\downarrow\rangle$ and then transmit right into the junction bulk
, where $|h\uparrow;e\downarrow\rangle$ indicates a state composed of 
a hole in the top layer bound to an electron in the bottom layer. 
However, single-particle tunneling destroys the excitons everywhere 
and leads to the attenuation of counterflow Josephson current in the bulk. 
When reversing the direction of external tunneling current [see Fig. \ref{Fig2}(c)], 
the roles of electrons and holes are exchanged and right-going but 
opposite polarized excitons $|e\uparrow;h\downarrow\rangle$ occur, 
where $|e\uparrow;h\downarrow\rangle$ indicates a state composed of 
an electron in the top layer bound to a hole in the bottom layer. 
Similarly, applying the $+\hat{z}$($-\hat{z}$)-direction tunneling current to the right edge 
will generate ``left"-going excitons $|h\uparrow;e\downarrow\rangle$ ($|e\uparrow;h\downarrow\rangle$) [see Figs. \ref{Fig2}(d)-(e)].
It turns out that the external tunneling currents with parallel (anti-parallel) polarity will 
inject counterflow Josephson current in the opposite (same) direction as shown in the insets of Figs. \ref{Fig2}(b)-(e). 
   
\subsection{Calculation of induced magnetic field due to excitonic Josephson effect}
We next consider a Corbino-geometry excitonic Josephson junction that can generate circular Josephson current [see Fig. \ref{Fig1}(c)].      
The Corbino can be divided into a set of rings with radius which ranges from $R_{\rm min}$ to $R_{\rm max}$. 
A single ring of the specific radius $r$ can be viewed as a bent Josephson junction with $L=2\pi r$.  
We firstly calculate the phase distribution for the junction of $L=2\pi R_{\rm{min}}$ by the LLG equation 
and then acquire the phase distribution for other values of $r$ by taking the azimuthal symmetry into account.
The Josephson current is similarly calculated by Eq. (\ref{Js}). 
By using the Biot-Savart Law, we finally obtain the induced magnetic field:   
\begin{widetext}
\begin{align}
B_{J}(z)=\frac{\mu_0\langle J_s(R_{\rm{min}},\theta)\rangle_{\theta} zdR_{\rm{min}}}{2} 
\left[\frac{1}{(R^2_{\rm{min}}+z^2)^{3/2}}-\frac{1}{(R^2_{\rm{max}}+z^2)^{3/2}}\right],  
\end{align} 
\end{widetext}
where $d$ is the interlayer separation, $z$ is the distance above the center of the bilayer, 
and $\langle \cdot\cdot\cdot\rangle_{\theta}$ is the average over the angular axis of polar coordinate. 

\subsection{Choice of units, identification of critical current and determination of parameters}
\label{unit}
Both two geometries we consider are discussed based on a length scale, namely, Josephson length: 
\begin{equation}
\lambda=\sqrt{2\rho_s/n\Delta_t}.  
\label{lambda}
\end{equation} 
Two units for Josephson current and tunneling current read 
$J_{s0}=e\rho_s/\hbar\lambda$ and $J_{t0}=en\Delta_t/2\hbar$ throughout this paper. 
We identify the critical interlayer tunneling current by finding the upper and lower
boundaries at which the junction departures from the coherent state, i.e., $m_z$ begins to become nonzero.  
The main focus of the present work is the typical quantum Hall bilayer of $\lambda\sim45\sf{\mu}$m ($\Delta_t=10^{-8}E_0$)\cite{Hyart:PRB2011}, 
which corresponds to the samples fabricated by Eisenstein's group\cite{Spielman:PRL2000}. 
Here the Coulomb interaction $E_0=e^2/\epsilon l_B$ serves as the energy scale and $E_0\sim 7$meV.      
The other parameters we use are listed as follows: $\beta=0.02E_0$ and $\rho_s=0.005E_0$, 
which were derived from the mean-field calculation\cite{Hyart:PRB2011}.  
  
\begin{figure} 
\centering
\vspace{0.0cm}
\begin{minipage}[t]{0.5\linewidth}
\centering
\includegraphics[width=4.7cm]{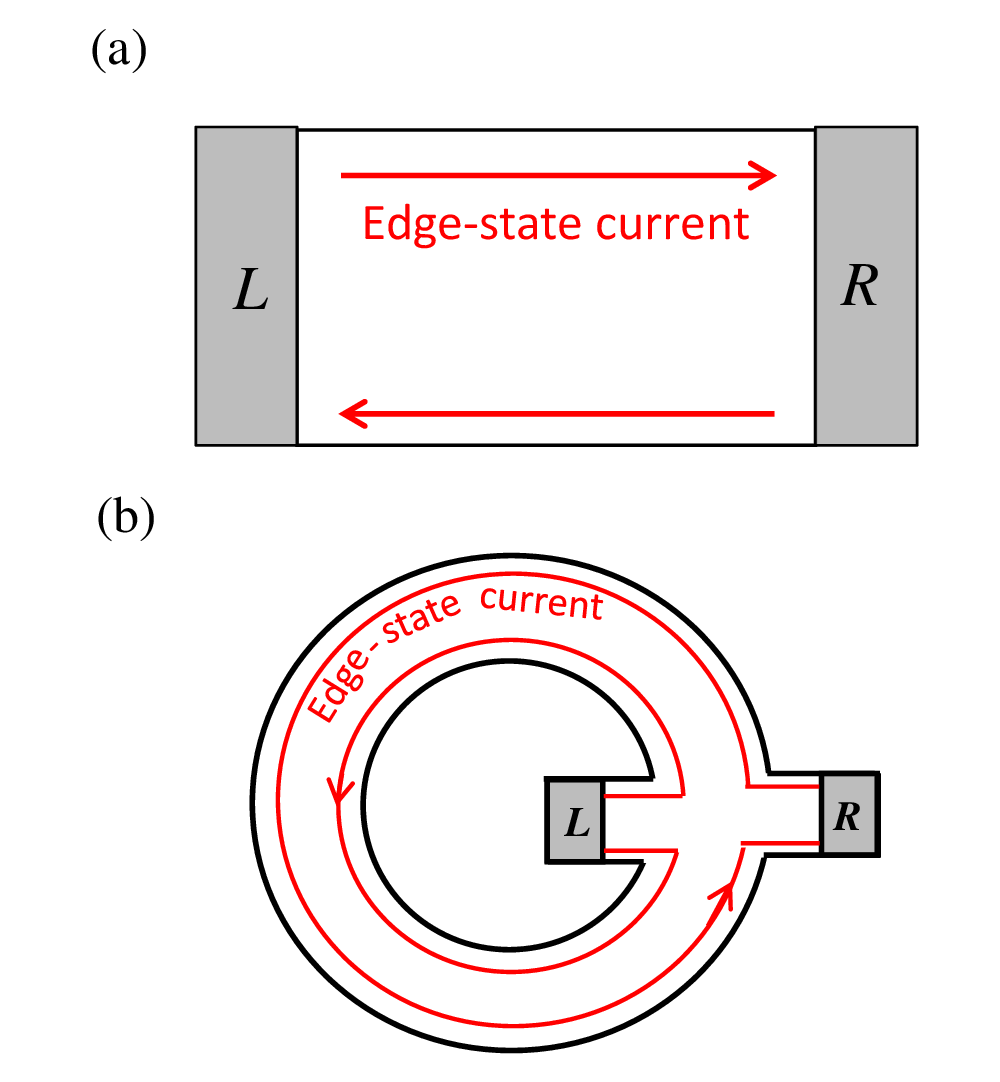}
\end{minipage}%
\begin{minipage}[t]{0.5\linewidth}
\centering
\includegraphics[width=4.0cm]{EXJETCLFig3b.eps}
\end{minipage}%
\vspace{0.3cm}
\begin{minipage}[t]{1.0\linewidth}
\centering
\includegraphics[width=5.0cm]{EXJETCLFig3c.eps}
\end{minipage}%
\caption{(color online) 
(a) and (b) depict the junction geometry: (a) the standard Hall bar geometry 
and (b) the Corbino geomtery. 
(c)-(e) summarize the key results of Huang's experiment (a realization of short Josephson junction): 
(c) Josephson-like $I$-$V$ characteristic without the compensating current applied.  
The inset shows the measurement configuration.     
(d) Josephson-like $I$-$V$ characteristic for different values of the compensating current. 
The numbers below the traces labels the corresponding value of the compensating current $I_{tL}/I_{t0}$.
The $I$-$V$ curves are offset by $(I_{tL}/2I_{t0})$mV. 
(e) The critical currents as a function of the compensating current. 
The measured current $I_{tL(R)}=J_{tL(R)}A$ and the unit $I_{t0}=1$nA, 
where $A$ labels the effective cross area of external tunneling currents.  
$A$ is difficult to determine through the existing experimental information.     
The data of (c)-(e) are reproduced from Ref. \cite{Huang:PRL2012}.  
}
\label{Fig3}
\end{figure}       
\section{Analysis of role of the compensating current}

\begin{figure} 
\includegraphics[width=7.5cm]{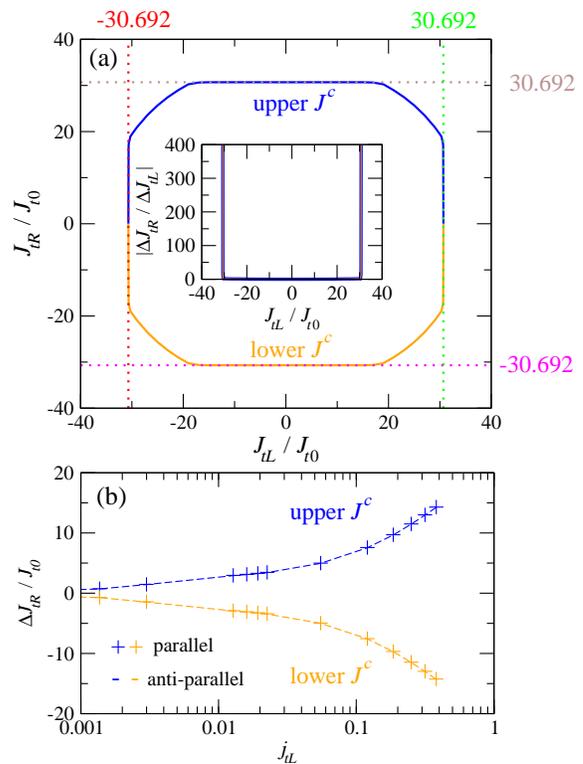}
\caption{(color online) (a) The calculated upper and lower critical values of 
the external tunneling current $J_{tR}$ as a function of its compensating current $J_{tL}$. 
The insert: the corresponding slopes $\Delta J_{tR}/\Delta J_{tL}$ as a function of $J_{tL}$. 
(b) The identification of critical exponents near two phase transition points occurring at $J_{tL}=\pm 30.692J_{t0}$. 
Here $\Delta J_{tR}=J^c(J_{tL})-J^c(\pm 30.692J_{t0})$ and $j_{tL}=|(J_{tL}-\pm 30.692J_{t0})/\pm 30.692J_{t0}|$. 
The choice of $\pm$ is up to which phase transition point we are discussing. 
By fitting to the numerical results presented in this figure, we extract the exponent $\gamma$, 
which is defined as $\Delta J_{tR}\propto j_{tL}^{\gamma}$, and find $\gamma\sim0.5$ for any phase transition point.   
} 
\label{Fig4}
\end{figure}

\begin{figure} 
\includegraphics[width=7.5cm]{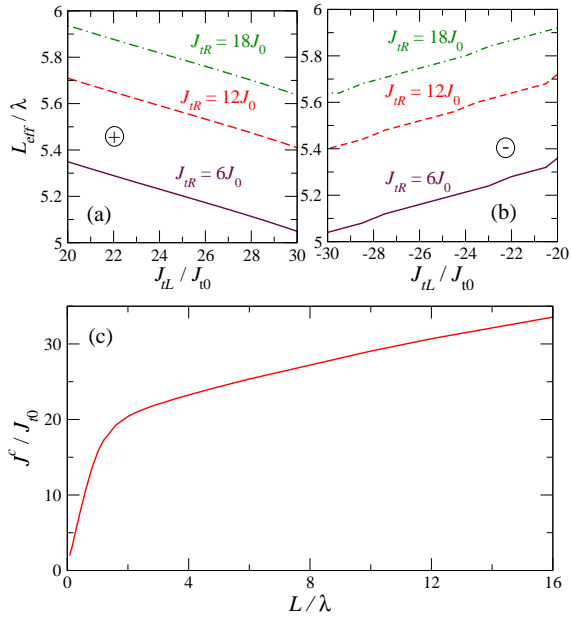}
\caption{(color online)  (a) and (b) are the effective length of the right EC/EC junction 
as a function of the corresponding compensating current $J_{tL}$ for the parallel polarity $\oplus$ 
and anti-parallel polarity $\ominus$ with the right tunneling current $J_{tR}=6,12,18J_{t0}$. 
(c) The junction-length dependence of critical current $J^c$ without the compensating current applied ($J_{tL}=0$). 
} 
\label{Fig5}
\end{figure}

Figs. \ref{Fig3}(a)-(b) show that edge-state currents inevitably contribute to 
the coupling of the left and right edges for the Hall-bar geometry 
while two edge-state currents separately flow along the inner and outer boundaries 
so as not to connect the left and right edges for the Corbino geometry\cite{Fink:PRL2011}. 
To avoid the contribution of edge-state currents, Fig. \ref{Fig3}(b) is the main setup we consider here 
and its corresponding junction length roughly approximates to the difference of the inner and outer radius.  
The realistic Corbino geometry possesses the junction length $L\sim0.54$mm\cite{Huang:PRL2012}and 
in the context of the typical quantum Hall bilayer\cite{Spielman:PRL2000} ($\lambda=45\mu$m),  
the junction length reads $L\sim12\lambda$. 
The large part of this paper therefore focuses on the case of $L=12\lambda$ later.        

\subsection{Nonequilibrium phase transition}
The realization of the short junction with $L=0.6\lambda$\cite{YFHsuPRL} --- 
Huang's experiment \cite{Huang:PRL2012}--- is devoted to analyzing Josephson-like behavior, 
in which the interlayer voltage suddenly emerges when applying tunneling current up to critical values: the upper and lower $I^c$ [see Fig. \ref{Fig3}(c)].  
They found the upper and lower critical currents are correlated with its compensating current 
--- the tunneling current exerted on the other edge and such coupling of the tunneling currents at two edges would disappear 
when $|I_{tL}|>16$nA [see Fig. \ref{Fig3}(d)]. 
The disappearance phenomenon will be discussed later in Sec. \ref{Josephsonbreakdown} 
and we focus on how the tunneling currents at two edges correlate with each other here.       
Huang's experiment quantifies this coupling through the plot of the critical currents as function of the compensating current[see Fig. \ref{Fig3}(e)].   
Therefore, we also display the similar plot for the long junction in Fig. \ref{Fig4} to analyze the role of the compensating current.  
Over a wide range of $J_{tL}$, the upper and lower critical currents nearly keep constant 
and are symmetric with respective to $J_{tR}=0$ [see Fig. \ref{Fig4}(a)]. 
Near $J_{tL}=\pm30.692J_{t0}$, however, the critical currents rapidly fall to zero.     
The sharp jump of critical currents $J^c$ indicates the right edge is switched 
from a superfluid to resistive state. 
The right edge undergoes a phase transition under the condition of compensating-current-driven nonequilibrium\cite{Nakamura:PRL2012,Matsumoto:PRD2018}. 
With slowly adjusting $J_{tL}$, it is identified as a first-order phase transition since $|J^c(J_{tL}=\pm30.692J_0)|=15.999J_0$ 
and $|J^c(J_{tL}=\pm30.6925J_0)|=0$ (The giant change in critical currents hints possible incontinuity). 
We furthermore define new critical exponents: 
\begin{equation}
  \Delta J_{tR}\propto \left\{
   \begin{array}{ll}
    (30.692J_{t0}-J_{tL})^{\gamma^+} & \mbox{for} J_{tL}\lesssim 30.692J_{t0},  \\
    (J_{tL}+30.692J_{t0})^{\gamma^-} & \mbox{for} J_{tL}\gtrsim -30.692J_{t0}, 
    \end{array} \right.
\end{equation}
where $\Delta J_{tR}=J^c(J_{tL})-J^c(\pm 30.692J_{t0})$. 
The fits to our numerical results extract the values of exponents [see Fig. \ref{Fig4}(b)]: 
$\gamma^+=0.4939$, $\gamma^-=0.4999$ for the upper $J^c$ curve. 
For the lower $J^c$ curve, the values of $\gamma^+$ and $\gamma^-$ 
are exactly exchanged because of electron-hole symmetry. 

\subsection{Junction-length reduction effect}
Why the compensating current can largely reduce the critical currents as $J_{tL}\approx\pm30.692J_{t0}$ 
even if the Josephson coupling is so weak? 
As have been illustrated in Fig. \ref{Fig1}(b), the long junction can be decomposed 
into two nearly independent EC/EC junctions. 
We here identify the breakpoint occurring at $J_s=0$ or where $J_s$ reaches its minimum 
and determine the effective length of the right EC/EC junction as shown in Figs. \ref{Fig5}(a)-(b). 
We find, regardless of the polarity, the compensating current $J_{tL}$ decreases 
the effective length of the right junction and hence leads to the jump of the critical currents. 
It is quite intuitive or shown in Fig. \ref{Fig5}(c) that the critical current would decrease 
with decreasing the junction length.

\section{other interesting prediction}
\subsection{Discussion on Josephson breakdown effect}

\label{Josephsonbreakdown}

\begin{figure} 
\includegraphics[width=6.8cm]{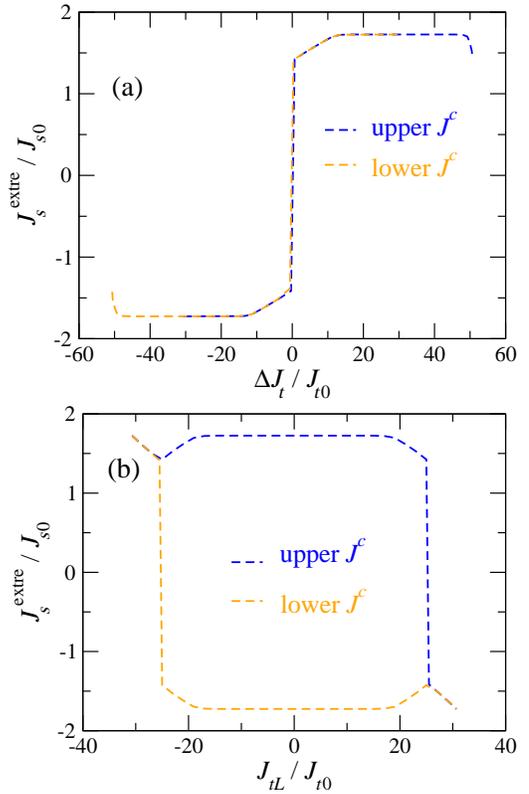}
\caption{(color online)  The spatial extrema of Josephson current $J_s^{\rm{extre}}$ as a function of (a) the difference of 
two tunneling currents $\Delta J_{t}=J_{tR}-J_{tL}$ 
and 
(b) the compensating current $J_{tL}$ for the upper and lower critical points of $J_{tR}$.  
} 
\label{Fig6}
\end{figure}
Now let us turn our attention to the disappearance phenomenon of the coupling of 
the two edge tunneling currents shown in Fig. \ref{Fig3}(b) occurring as $|I_{tL}|>16$nA. 
For this disappearance phenomenon, the main body of Ref. \cite{Huang:PRL2012} furthermore demonstrates that 
it is accompanied with the occurrence of the interedge voltage.  
Ref. \cite{YFHsuPRL} has attributed this phenomenon to the breakdown of Josephson effect---
when Josephson current attains some critical value, the Josephson effect would collapse 
and the external tunneling currents will prefer to converting into edge-state currents.  
We here comment on whether this breakdown effect occurs also in the long junction or not. 
Differing from the short junction, the upper and lower $J^c$ curves are 
always symmetric with respect to $J_{tR}=0$ as if the Josephson breakdown effect already happens 
and the applied compensating current is limited to a range of $J_{tL}=-30.692J_{t0}\sim30.692J_{t0}$ 
beyond which coherent interlayer tunneling disappears [see Fig. \ref{Fig4}(a)]. 
We have performed numerical calculation demonstrating that over the range of $J_{tL}=-30.692J_{t0}\sim30.692J_{t0}$, 
static solutions can exist and there was not found any critical variation. 
We therefore believe that the breakdown effect does not occur in the long junction. 

We furthermore give more detail analysis through Fig. \ref{Fig6}. 
The difference of external tunneling currents $\Delta J_t$ plays the similar role 
as the relative phase in the conventional Josephson junction\cite{Golubov} 
while it is easier to compare with the experiment directly based on the compensating current $J_{tL}$. 
In Fig.  \ref{Fig6}, we therefore plot the spatial extrema of Josephson current $J_s^{\rm extre}$  
as a function of not only $\Delta J_t$ but also $J_{tL}$. 
We find that $J_{s}^{\rm{extre}}$ rises or drops to saturation 
over the range of $\Delta J_t=20J_{t0}\sim40J_{t0}$ or $\Delta J_t=-20J_{t0}\sim-40J_{t0}$ [see Fig. \ref{Fig6}(a)], 
which corresponds to $J_{tL}=-20J_{t0}\sim20J_{t0}$ [see Fig. \ref{Fig6}(b)].   
With increasing the compensating current, if the Josephson-breakdown regime is achieved,   
it necessarily occurs at $J_{tL}=-20J_{t0}\sim20J_{t0}$ where the $J^c$ curves hold horizontal [see Fig. \ref{Fig4}(a)].  
Measuring the interedge voltage will help us clarify the junction being in 
the weak Josephson coupling regime or Josephson-breakdown regime. 
Alternatively, after increasing the compensating current beyond $\pm 20J_{to}$, 
$|J^c|$ begins to fall [see Fig. \ref{Fig4}(a)], providing an unique signature 
for the weakly Josephson coupling, namely, Josephson fall. 

\subsection{The crossover behavior with varying junction length}

\begin{figure} 
\includegraphics[width=8.5cm]{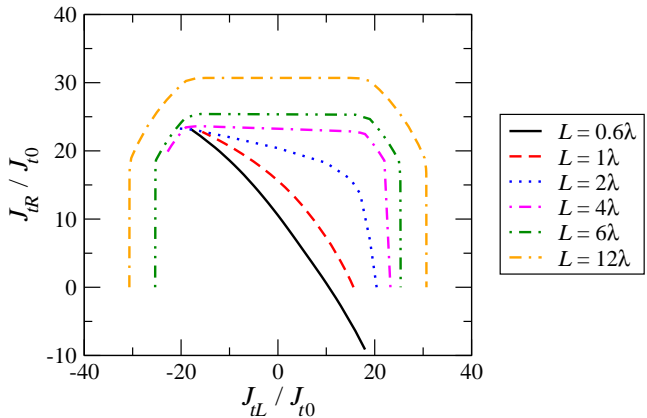}
\caption{(color online) The critical value of the external tunneling current $J_{tR}$ versus 
the compensating current $J_{tL}$ for different junction length $L$.
} 
\label{Fig7}
\end{figure}

Since the dependence of the critical currents on the compensating current is so distinct for the short and long junctions,  
we next want to understand the crossover behavior with increasing junction length through Fig. \ref{Fig7}.  
Because the lower $J^c$ curve can be produced through doing 
the electron-hole transformation: $J_{tR}\rightarrow-J_{tR}$, $J_{tL}\rightarrow-J_{tL}$ on the upper $J^c$ curve, 
in Fig. \ref{Fig7}, we display only the upper $J^c$ curve for conciseness. 
Fig. \ref{Fig7} shows that, with increasing the junction length, the curve is gradually skew and no abrupt change occurs. 
Moreover, the Josephson fall already can be found as $L=4\lambda$ 
while the weakly ``symmetric" Josephson regime can be achieved as $L\sim5\lambda$.  
The values of $5\lambda$ happens to meet the junction length for the typical quantum Hall bilayer\cite{Spielman:PRL2000} of Hall-bar geometry 
($L\sim 225\mu$m) but the Hall-bar geometry may be difficult to coincide with our calculation due to the influence of edge-state current. 
Replacing the usually-used side electrodes with the top and back electrodes would be a method to 
avoid edge-state currents although it is a big technological challenge.       
  
\subsection{The induced magnetic field due to Josephson current in a Corbino geometry} 

\begin{figure} 
\includegraphics[width=7.0cm]{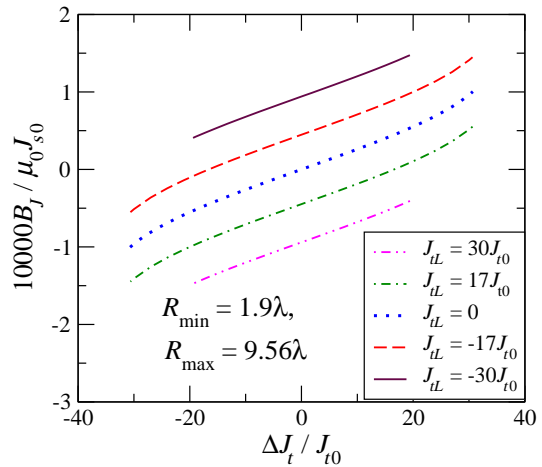}
\caption{(color online) 
The induced magnetic field $B_{J}$ due to circular Josephson current of 
a Corbino-geometry excitonic Josephson junction at $z=2.22\lambda$ as a function of 
the difference of two external tunneling currents $\Delta J_{t}=J_{tR}-J_{tL}$
for $R_{\rm{min}}=1.9\lambda$ and $R_{\rm{max}}=9.56\lambda$. 
The curves are offset by the corresponding $J_{tL}$.   
The interlayer separation $d=1.6\ell_B$, where $\ell_B$ is the magnetic length.} 
\label{Fig8}
\end{figure}

Next Fig. \ref{Fig8} shows the results for the Corbino-geometry excitonic Josephson junction, 
which is depicted in Fig. \ref{Fig1}(c) (the curves is offset by the corresponding compensating current 
for clarity and a without-offset version is given in Appendix A).   
In Fig. \ref{Fig8}, except for the minimum radius $R_{\rm{min}}$, 
the other parameters are determined according to the realistic situation of experiments. 
The minimum radius for the typical Corbino is roughly 0.16mm or equivalently $R_{\rm{min}}\sim3.56\lambda$ 
instead of $R_{\rm{min}}=1.9\lambda$ that we choose for increasing the numerical efficiency. 
But, the investigated Corbino of $\lambda<2\pi R_{\rm{min}}<2\pi R_{\rm{max}}$ can already capture the physics 
of the long junction to a qualitative level and such a Corbino with smaller $R_{\rm{min}}$ is easily realized by etching. 
We find, differing from the short junction\cite{YFHsuPRL}, the dependence of the induced magnetic field $B_J$ on the difference 
of two tunneling currents $\Delta J_t$ can have apparent curvature. 
The curve however becomes linear when $J_{tL}$ reaches $\pm30J_{t0}$. 
This is because $J_{tL}$ decreases the effective length of the EC/EC junction on the opposite side 
and drives the investigated Corbino into the short-junction regime of a linear dependence\cite{YFHsuPRL}.    
Moreover, the extremely subtle magnetic field must be measured by the scanning superconducting quantum interference device (SQUID). 
To our best knowledge, the resolution of the typical scanning SQUID is up to $\sim10$pT 
at a sensor-to-sample distance of $\sim 100$nm and the current technology even improves the resolution to $\sim1$pT\cite{Oda}.  
We estimate $B_J$ on the scale $\sim100$pt and it is measurable without doubt.       

\section{conclusion} 
In conclusion, we predict a nonequilibrium phase transition occurring in the long junction of weak Josephson coupling 
and find the effective length reduction effect of the compensating current.  
The sample size is not highly tunable in experimental measurement and therefore 
this length reduction effect will be largely helpful in observing 
the interesting crossover behavior predicted in Ref. \cite{YFHsu:NJP2018}. 
We furthermore discuss the possibility of the breakdown of Josephson effect 
and suggest measuring the interedge voltage and Josephson fall\cite{note2} 
to distinguish the Josephson breakdown effect from weak Josephson coupling. 
We also calculate the induced magnetic field in the Corbino-geometry Josephson junction 
to suggest the detection of Josephson current. 
It should be noted that there are still very much theoretical effort called for, 
such as developing Bogolubov-deGennes description, 
exactly identifying phase transition (especially for it being first-order or second-order), 
systematically exploring the Josephson breakdown effect and etc. 
We believe the present work together with Ref. \cite{YFHsuPRL}--- 
excitonic Josephson effect induced by interlayer tunneling current ---will bring new attention 
to the condensed matter physics community.  

\begin{figure} 
\vspace{0.5cm}
\includegraphics[width=7.0cm]{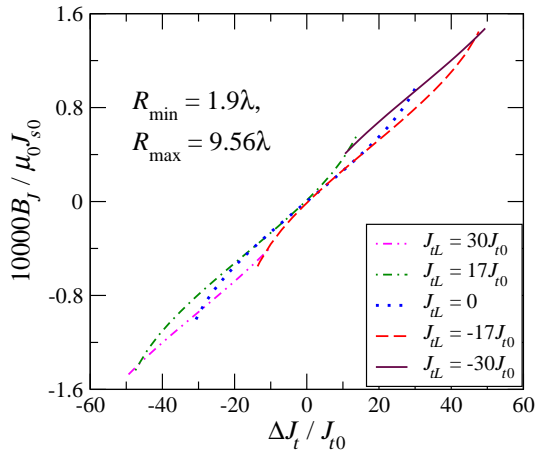}
\setlength{\belowcaptionskip}{-0.5cm} 
\caption{(color online) The without-offset version for Fig. \ref{Fig8}, where $B_J$ and $\Delta J_{t}$ denote the induced magnetic field 
and the difference of two external tunneling currents, respectively.
}  
\label{FigA1}
\end{figure} 
\section*{Acknowledge}
We are grateful to  W. Dietsche, A. H. MacDonald, B. Rosenstein, Jheng-Cyuan Lin, Sing-Lin Wu and Chien-Ming Tu for valuable discussion. 
This work were financially supported by Ministry of Science and Technology and by National Center for Theoretical Sciences of Taiwan.
\section*{Appendix A: The without-offset version for Fig. \ref{Fig8}}
In Fig. \ref{FigA1}, we display the original curves of Fig. \ref{Fig8}, 
being not offset, to capture more definite understanding for the $\Delta J_t$ dependence. 
Similar to the short junction discussed in Ref. \cite{YFHsuPRL}, the curves for different $J_{tL}$ approaches each other  
but apparent derivation exists for large $\Delta J_t$.  
That is to say, for the long junction, the magnitude of the induced magnetic is dependent on not only the difference of 
two external edge tunneling currents but also their individual values, 
which can be regarded as a characteristic of weak Josephson coupling.    
This is because in the weakly Josephson-coupled regime, the edge property becomes prominent.


\begin{thebibliography}{50}
\bibitem{note1} Josephson length is a well-known characteristic length of quantum Hall bilayer exciton condenstates. 
In this paper, its definition is given in Sec. \ref{method}.
\bibitem{Js} This inference is based on that supercurrent is proportional to the slope of the condensate phase.
\bibitem{note2} The definition can be found in Sec. \ref{Josephsonbreakdown}.
\end{thebibliography}
\end{document}